\documentclass{pasj01}
\newcommand{\jcap}{J. Cosmol. Astropart. Phys.}
\Received{$\langle$reception date$\rangle$}
\Accepted{$\langle$acception date$\rangle$}
\Published{$\langle$publication date$\rangle$}

\begin{document}

\title{Heavy axion-like particles and MeV decay photons from nearby type Ia supernovae}
\author{Kanji Mori}%
\altaffiltext{}{Research Institute of Stellar Explosive Phenomena, Fukuoka University, 8-19-1 Nanakuma, Jonan-ku, Fukuoka-shi, Fukuoka 814-0180, Japan}
\email{kanji.mori@fukuoka-u.ac.jp}


\maketitle

\begin{abstract}
Axion-like particles (ALPs) are hypothetical  bosons which may couple with photons. Since many ALPs can be emitted from hot and dense astrophysical plasma, nearby supernovae (SNe) are a possible probe into their properties including the ALP mass $m_a$ and the coupling constant $g_{a\gamma}$ between ALPs and photons. I calculated ALP emission from a type Ia SN (SN Ia) model with the near-Chandrasekhar mass. It is found that the ALP luminosity from  SNe Ia reaches $\sim10^{43}(g_{a\gamma}/10^{-10}\; \mathrm{GeV}^{-1})^{2}$ erg s$^{-1}$ if $m_a\lesssim1$ MeV. Heavy ALPs emitted from SNe are unstable and decay into photons. I predict the time delay and the flux of decay photons that reach Earth from a nearby SN Ia. It is found that the decay photons may provide a constraint on $g_{a\gamma}$ which is as stringent as an SN 1987A limit if an SN Ia is located 1 kpc away or closer and next-generation MeV $\gamma$-ray satellites observe it $\sim1-10$ years after the explosion.
\end{abstract}

\section{Introduction}
Axion-like particles (ALPs) are a class of hypothetical bosons beyond the standard model of particle physics (e.g. \cite{2020arXiv201205029C,2020PhR...870....1D}). Axions are pseudo Nambu-Goldstone bosons originally introduced to solve the strong CP problem of quantum chromodynamics \citep{1978PhRvL..40..223W,1978PhRvL..40..279W}. ALPs, which share similar properties with axions, emerge as a result of string theory compactifications \citep{2006JHEP...06..051S,2010PhRvD..81l3530A}. It is notable that particle models with ALPs may solve the electroweak hierarchy problem \citep{2015PhRvL.115v1801G}.

The interaction between an ALP and photons is described by the Lagrangian \citep{1988PhRvD..37.1237R}
\begin{equation}
\mathcal{L}_{a\gamma\gamma}=-\frac{1}{4}g_{a\gamma}a\tilde{F}^{\mu\nu}F_{\mu\nu},
\end{equation}
where $g_{a\gamma}$ is the coupling constant, $a$ is the ALP field, $F_{\mu\nu}$ is the electromagnetic tensor, and $\tilde{F}_{\mu\nu}$ is its dual. The mass $m_a$ and the coupling $g_{a\gamma}$ of heavy ALPs with $m_a\gtrsim10$ keV have been explored by beam dump experiments \citep{2016PhLB..753..482J,2017JHEP...12..094D,2019JHEP...05..213D,2020PhRvL.125h1801B} and astronomical objects including globular clusters \citep{2020PhLB..80935709C}, white dwarfs \citep{2021arXiv210200379D}, and the neutrino signal \citep{2018arXiv180810136L,2020JCAP...12..008L} and $\gamma$-rays \citep{2011JCAP...01..015G,2018PhRvD..98e5032J} from SN 1987A. Also, implications of heavy ALPs on cosmological phenomena including Big Bang nucleosynthesis and the cosmic microwave background have been discussed \citep{2011JCAP...02..003C,2020JCAP...05..009D}.

Since many ALPs may be produced in hot plasma in supernovae (SNe) depending on $m_a$ and $g_{a\gamma}$, a nearby SN can be used as a probe of ALPs. Although core-collapse SNe produce more ALPs than type Ia SNe (SNe Ia) do because of the higher temperature, the next nearby SN after SN 1987A can be type Ia \citep{2011MNRAS.412.1473L}. It is therefore desirable to predict the ALP emission from SNe Ia to discuss a possible constraint that can be obtained from the event.

A heavy ALP emitted from an SN decays into two photons during propagation to the Solar System. It is hence possible to constrain the ALP parameters by $\gamma$-ray observations of a nearby astronomical object which emits a lot of ALPs. In this paper, I calculate the ALP production in a nearby SN Ia and predict the $\gamma$-ray flux which originates from the ALP decay.

This {paper} is organized as follows. In Section 2, I describe the prescription of the ALP production and decay and the SN Ia model. In Section 3, the intrinsic ALP emission from the SN Ia is shown. In Section 4, the $\gamma$-ray flux on Earth is estimated and its observability is discussed. In Section 5, the results are summarized and future prospects are explained.

\section{Method}
In this Section, I describe the ALP production and decay rates and the SN Ia model adopted in this work.
\subsection{ALP production}
In this study,  we consider the Primakoff process \citep{1951PhRv...81..899P}
\begin{eqnarray}
\gamma+e^-&\rightarrow& a+e^-\\
\gamma+Ze&\rightarrow& a+Ze
\end{eqnarray}
induced by electrons and ions and photon coalescence
\begin{eqnarray}
\gamma+\gamma\rightarrow a
\end{eqnarray}
as the ALP production processes. 

The Primakoff rate is given by \citep{2000PhRvD..62l5011D,2020PhLB..80935709C,2020JCAP...12..008L}
\begin{eqnarray}
\Gamma_{\gamma\rightarrow a}=g_{a\gamma}^2\frac{T\kappa^2}{32\pi}\frac{p}{E}\left(\frac{((k+p)^2+\kappa^2)((k-p)^2+\kappa^2)}{4kp\kappa^2}\right.\nonumber\\
\left.\times\ln\left(\frac{(k+p)^2+\kappa^2}{(k-p)^2+\kappa^2}\right)
-\frac{(k^2-p^2)^2}{4kp\kappa^2}\ln\left(\frac{(k+p)^2}{(k-p)^2}\right)-1\right),\label{gamma}
\end{eqnarray}
where $p=\sqrt{E_a^2-m_a^2}$ and $k=\sqrt{\omega^2-\omega_\mathrm{pl}^2}$ are the momenta and $E_a$ and $\omega$ are the energies of ALPs and photons, respectively. The energy conservation leads to a relation $E_a=\omega$. The plasma frequency is denoted by $\omega_\mathrm{pl}$ and
\begin{eqnarray}
\kappa=\sqrt{\frac{4\pi\alpha n^\mathrm{eff}}{T}}
\end{eqnarray}
is the inverse Debye-H\"uckel length, where $\alpha$ is the fine structure constant and
$n^\mathrm{eff}=n_e^\mathrm{eff}+n^\mathrm{eff}_\mathrm{ion}$
is the sum of the effective electron and ion number densities. The plasma frequency is evaluated as (Chapter 6 and Appendix D in \cite{1996slfp.book.....R})
\begin{eqnarray}
    \omega_\mathrm{pl}=\sqrt{\frac{4\pi\alpha n_e}{E_\mathrm{F}}}\approx28.7\; \mathrm{eV}\frac{(Y_e\rho)^\frac{1}{2}}{(1+(1.019\times10^{-6}Y_e\rho)^\frac{2}{3})^\frac{1}{4}},
\end{eqnarray}
where $n_e$ is the electron number density, $E_\mathrm{F}$ is the electron Fermi energy, $Y_e$ is the electron mole fraction, and $\rho$ is the density in units of g cm$^{-3}$. Equation (7) is an asymptotic formula for degenerate plasma and holds for both non-relativistic ($p_\mathrm{F}\ll m_e$) and relativistic ($p_\mathrm{F}\gg m_e$) cases, where $m_e$ is the electron mass and $p_\mathrm{F}$ is the electron Fermi momentum. Since the gas is always degenerate in the region of interest, $\omega_\mathrm{pl}$ is reasonably approximated by the equation.

The effective ion density is defined as
\begin{eqnarray}
    n^\mathrm{eff}_\mathrm{ion}=\frac{\rho}{m_\mathrm{u}}\sum_j\frac{Z_j^2X_j}{A_j},
\end{eqnarray}
where $\rho$ is the density, $m_\mathrm{u}$ is the atomic mass unit, and $Z_j$, $X_j$, and $A_j$ are the atomic number, the mass fraction, and the mass number of the $j$-th nucleus, respectively.

The effective electron density $n_e^\mathrm{eff}$ takes degeneracy into account. It is calculated as $n_e^\mathrm{eff}=R_\mathrm{deg}n_e$, where $n_e=\rho Y_e/m_\mathrm{u}$,  and $R_\mathrm{deg}$ is a coefficient defined as follows \citep{2015JCAP...02..006P,2019ApJ...881..158S}. We introduce a parameter $z=\rho Y_e/T^\frac{3}{2}$ where $T$ is the temperature. When $z$ with $T$ in K and $\rho$ in g cm$^{-3}$ is less than or equal to $5.45\times10^{-11}$, $R_\mathrm{deg}=1$. When $z\in(5.45\times10^{-11},\;7.2\times10^{-8}]$,
\begin{eqnarray}
R_\mathrm{deg}=0.63+0.3\arctan\left(0.65-9316z^{0.48}+\frac{0.019}{z^{0.212}}\right).
\end{eqnarray}
Finally, when $z>7.2\times10^{-8}$, 
\begin{eqnarray}
R_\mathrm{deg}=4.78\times10^{-6}z^{-0.667}.
\end{eqnarray}

The ALP production rate per a unit volume and time is then written as
\begin{eqnarray}
Q_a&=&2\int\frac{d^3\mathbf{k}}{(2\pi)^3}\Gamma_{\gamma\rightarrow a}\omega f(\omega)\nonumber\\
&=&\frac{1}{\pi^2}\int^\infty_{m_a}d\omega\omega^2\sqrt{\omega^2-\omega_\mathrm{pl}^2}\Gamma_{\gamma\rightarrow a}f(\omega),\label{prim}
\end{eqnarray}
where $f(\omega)$ is the photon Bose-Einstein distribution. The factor 2 in the second term stems from the degrees of freedom of photons.

The ALP production rate due to the photon coalescence is given by \citep{2000PhRvD..62l5011D,2020PhLB..80935709C,2020JCAP...12..008L}
\begin{equation}
\frac{d^2n_a}{dtdE}=g_{a\gamma}^2\frac{m_a^4}{128\pi^3}p\left(1-\frac{4\omega_\mathrm{pl}^2}{m_a^2}\right)^\frac{3}{2}e^{-\frac{E_a}{T}},\label{pcrate}
\end{equation}
and the ALP production rate for this reaction is then
\begin{equation}
Q_a=\int^\infty_{m_a}dE_aE_a\frac{d^2n_a}{dtdE_a}.\label{pc}
\end{equation}
It should be noted that this reaction is possible only when the kinematic condition $m_a>2\omega_\mathrm{pl}$ is satisfied.

When the ALP kinetic energy is smaller than the gravitational energy, the particle is trapped \citep{2003PhRvD..68e5004D,2020JCAP...12..008L}. In this case, the decay photons cannot be observed. However, the gravitational energy $E_\mathrm{G}=GM m_a/r$ between the star and an ALP is only $\sim7$ keV, where $G$ is the gravitational constant, $M\sim1.4M_\odot$ is the total mass of the star, and $r$ is the stellar radius. I here assumed $r\sim3\times10^3$ km. Since the gravitational energy is much less than ALP energies ($\sim$ MeV), the trapping effect is negligible.
\subsection{ALP decay}
Heavy ALPs are unstable particles. The mean free path (MFP) due to the ALP decay ($a\rightarrow \gamma\gamma$) is \citep{2018PhRvD..98e5032J,2020JCAP...12..008L}
\begin{eqnarray}
\lambda_{a\rightarrow \gamma\gamma}(E)=\frac{\beta\gamma}{\Gamma_{a\rightarrow \gamma\gamma}}
\end{eqnarray}
where $\gamma$ is the Lorentz factor of ALPs, $\beta=\sqrt{1-\gamma^{-2}}$, and
\begin{equation}
\Gamma_{a\rightarrow \gamma\gamma}=g_{a\gamma}^2\frac{m_a^3}{64\pi}.
\end{equation}
The MFP is then evaluated as
\begin{equation}
    \lambda_{a\rightarrow \gamma\gamma}\sim0.4\;\mathrm{ly}\;g_{10}^{-2}\left(\frac{E_a}{1\;\mathrm{MeV}}\right)\left(\frac{m_a}{1\;\mathrm{MeV}}\right)^{-4},
\end{equation}
where $g_{10}=g_{a\gamma}/(10^{-10}\;\mathrm{GeV}^{-1})$.
\subsection{SN Ia model}
\begin{figure}
 \begin{center}
  \includegraphics[width=80mm]{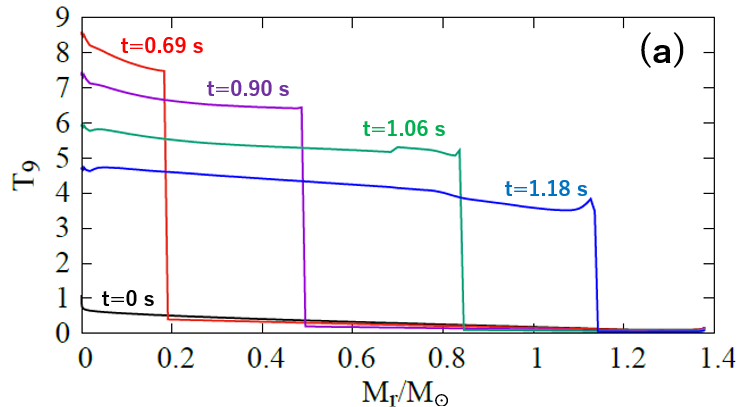}
    \includegraphics[width=80mm]{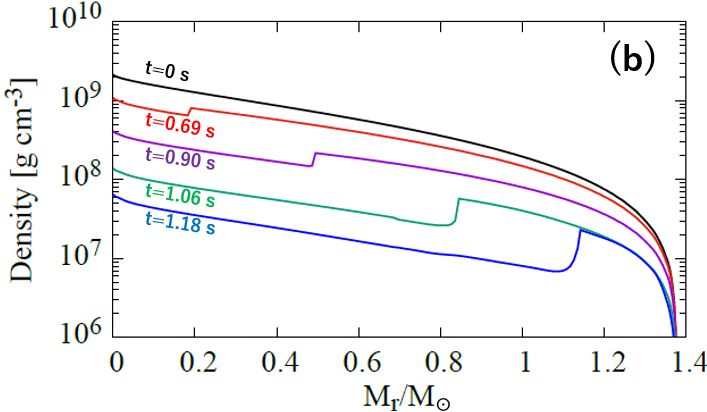}
    \includegraphics[width=80mm]{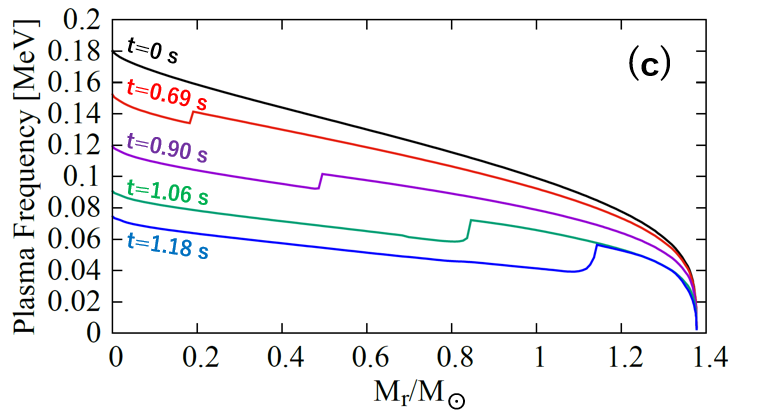}
 \end{center}
 \caption{{The profiles of (a) temperature, (b) density, and (c) plasma frequency in} the SN Ia model as a function of the mass coordinate $M_r$.}\label{W7}
\end{figure}
SNe Ia are a thermonuclear explosion of a carbon-oxygen white dwarf (WD). In near-Chandrasekhar mass models of SNe Ia, accretion occurs on the WD and its mass becomes close to the Chandrasekhar mass. As a result, carbon fusion is ignited at the center and the whole star becomes unbound. In this study, we adopt W7 \citep{1984ApJ...286..644N} as the SN Ia model, which is a one-dimensional model with pure deflagration. 

Figure \ref{W7} shows {the profiles of the temperature, the density, and the plasma frequency in the W7 model. Here $M_r$ is the mass coordinate defined as
\begin{eqnarray}
    M_r=4\pi\int^r_0\rho(r') r'^2dr'
\end{eqnarray}
and} $T_9$ is the temperature in units of $10^9$ K. The deflagration wave starts at the center and propagates toward the surface. When the flame passes, the material is heated and the temperature rises. The initial density is as high as $\sim10^9$ g cm$^{-3}$ because the star is close to the Chandrasekhar mass. Since the WD expands as the star explodes, the density decreases as a function of time. 

The Primakoff rate shown in equation (\ref{gamma}) is dependent on the chemical composition. In this calculation, I adopt the nuclear statistical equilibrium (NSE) prescription \citep{2009ADNDT..95...96S} when $T_9\geq3$ {and carbon-oxygen-neon plasma with $X(^{12}\mathrm{C})=0.475$, $X(^{16}\mathrm{O})=0.500$, and $X(^{22}\mathrm{Ne})=0.015$ when $T_9<3$. The NSE is a good approximation for hot plasma with $T_9\geq3$ where the majority of ALPs is produced. Because the NSE approximation is not valid when $T_9<3$, the composition is switched to the carbon-oxygen-neon plasma for cooler gas.  The NSE table shown by \citet{2009ADNDT..95...96S} considers 443 nuclei from neutron to krypton. All of these nuclei are considered in the calculation of the Primakoff rate.}  In the NSE, the nuclear abundances are determined only by $T$, $\rho$, and $Y_e$. I adopt a constant electron fraction $Y_e=0.4575$ on the basis of a network calculation \citep{2020ApJ...904...29M}.
\section{Intrinsic ALP production in the SN}
\begin{figure}
 \begin{center}
  \includegraphics[width=80mm]{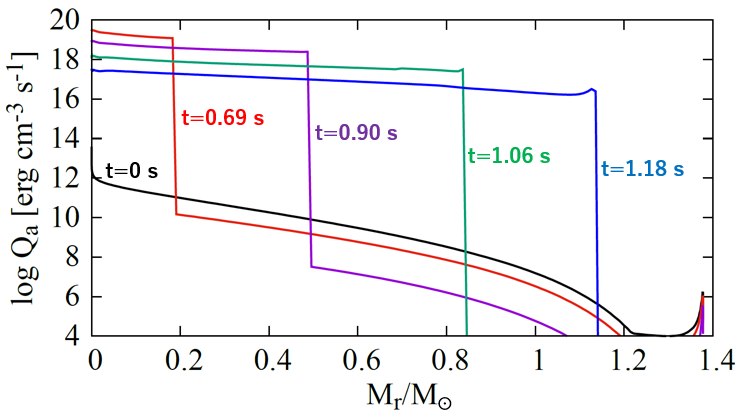}
 \end{center}
 \caption{The ALP production rate  per a unit volume and time as a function of the mass coordinate. The coupling constant and the ALP mass are fixed to $g_{10}=1$ and $m_a=0.1$ MeV, respectively.}\label{Qa}
\end{figure}
\begin{figure}
 \begin{center}
  \includegraphics[width=80mm]{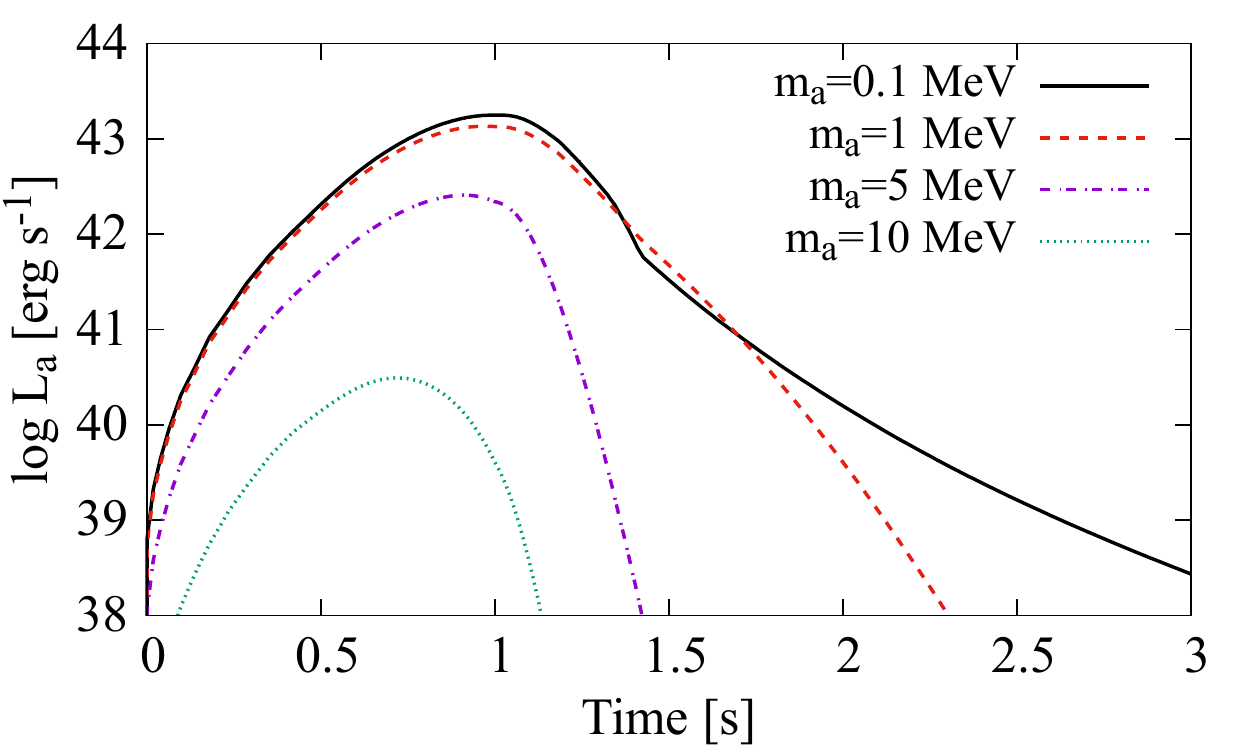}
 \end{center}
 \caption{The ALP luminosity as a function of time. The coupling constant is fixed to $g_{10}=1$.}\label{La}
\end{figure}
Using the SN model shown in figure \ref{W7} and equation (\ref{prim}), we can calculate the ALP emissivity from an SN Ia. Figure \ref{Qa} shows the profile of the ALP emissivity $Q_a$ defined in equations (\ref{prim}) and (\ref{pc}). In this figure, the ALP mass is fixed to $m_a=0.1$ MeV and the coupling constant is fixed to $g_{10}=1$. Since $Q_a$ rapidly increases as a function of $T$, ALPs are emitted behind the flame. Although $Q_a$ becomes smaller as the SN cools, the region where ALPs are emitted becomes wider because of the propagation of the flame.

In figure \ref{La}, I show the ALP luminosity
\begin{equation}
    L_a=4\pi\int Q_a(r)r^2dr,
\end{equation}
where the integral is performed from the center to the surface of the star. I adopt $g_{10}=1$ in this plot. It is seen that $L_a$ reaches its peak at $0.5-1$ s after the ignition. When $m_a\lesssim1$ MeV, $L_a$ reaches $\sim10^{43}$ erg s$^{-1}$ at its peak. Coincidentally, this value is close to peak photon luminosities of a typical SN Ia.  The ALP luminosity is much smaller when $m_a\gtrsim5$ MeV because the thermal photons are not energetic enough to  produce such heavy ALPs.

{In the case of $m_a=0.1$ MeV, we can see a bend in $L_a$ at $t\sim1.4$ s. This is an artifact that originates from a sudden switch from the NSE to the carbon-oxygen-neon plasma and will disappear if dynamical nucleosynthetic processes are considered.}
\begin{figure}
 \begin{center}
  \includegraphics[width=80mm]{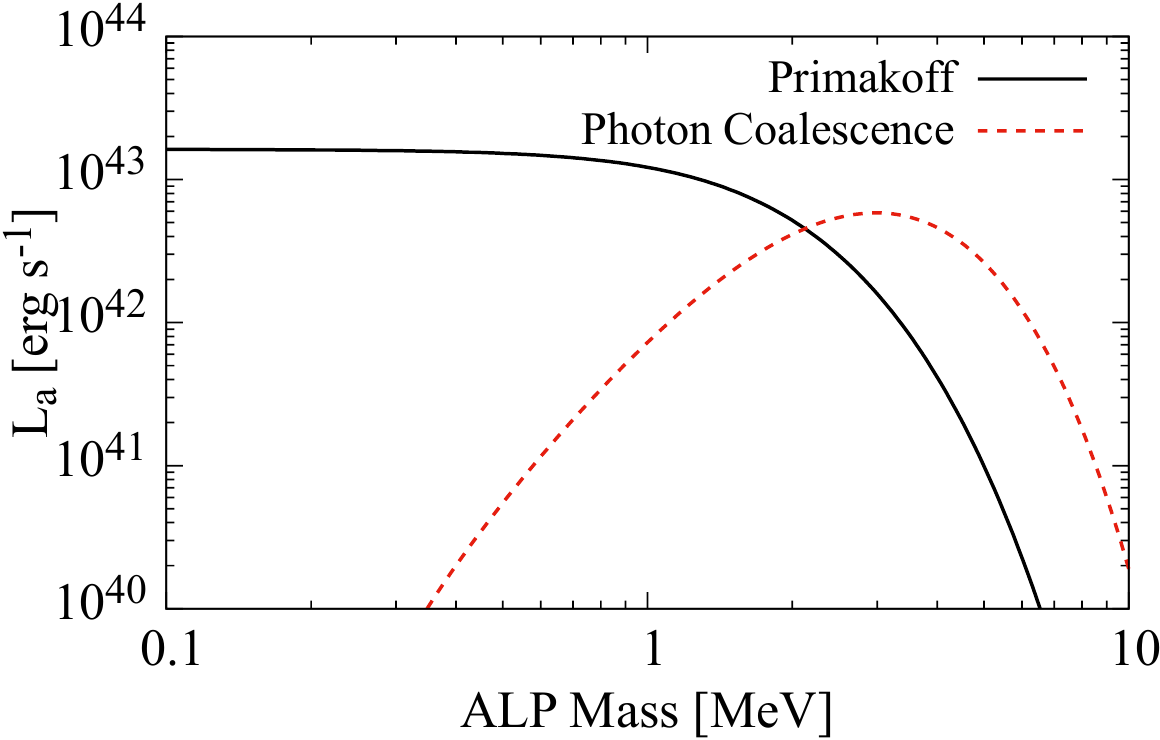}
 \end{center}
 \caption{The contribution of the Primakoff process and photon coalescence on the ALP luminosity as a function of the ALP mass. The result at $t=0.90$ s after the ignition is shown. The coupling constant is fixed to $g_{10}=1$.}\label{La_mass}
\end{figure}

{Figure \ref{La} is the total ALP luminosity induced by the Primakoff process and photon coalescence. Figure \ref{La_mass} shows the contribution of each process at $t=0.90$ s as a function of the ALP mass. It is seen that the Primakoff process is dominant if $m_a\lesssim 2$ MeV while photon coalescence is dominant if $m_a\gtrsim2$ MeV.}

In figures \ref{Qa}, \ref{La} and \ref{La_mass}, $g_{10}=1$ is assumed. However, it is notable that $Q_a$ and $L_a$ are simply proportional to $g_{10}^2$  as we can see from  equations (\ref{gamma}) and (\ref{pcrate}).
\section{Decay photons observed on Earth}
As described in Section 2.2, a heavy ALP can decay into two photons, which can be detected on Earth by $\gamma$-ray telescopes. In this Section, I predict the expected number and the flux of the decay photons.

\begin{figure*}
 \begin{center}
  \includegraphics[width=120mm]{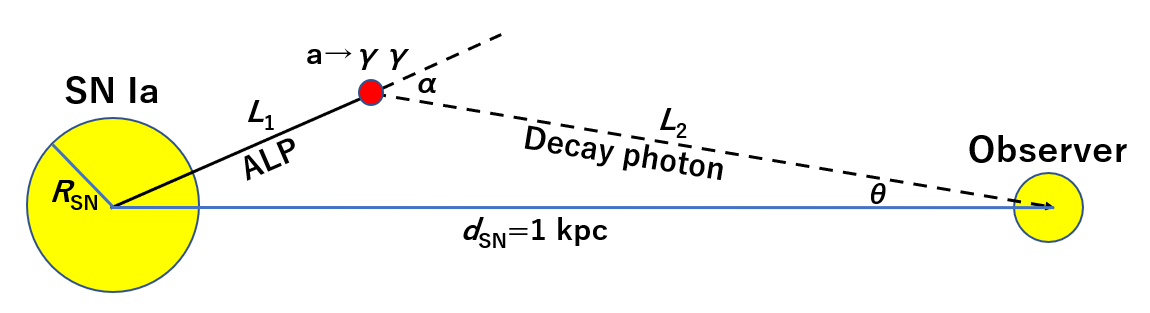}
 \end{center}
 \caption{A schematic picture of geometry between a nearby SN and Earth.}\label{sche}
\end{figure*}
The configuration discussed here is shown in figure \ref{sche}. In this Section, we assume that the distance to the SN Ia is $d_\mathrm{SN}=1$ kpc if not stated otherwise. The distance which an ALP travels before its decay is denoted by $L_1$ and the distance which the decay photon travels is denoted by $L_2$. The angle between the paths of the ALP and the photon is denoted by $\alpha$. It is easily shown that
\begin{equation}
    L_2=-L_1\cos\alpha+\sqrt{d_\mathrm{SN}^2-L_1^2\sin^2\alpha}.
\end{equation}

The angular distribution of decay photons is isotropic if an ALP is in its rest frame. However, since the SN ALP travels at a relativistic velocity, the decay photons are beamed toward the forward direction in our frame. The most likely angle is $\alpha\sim\gamma^{-1}$ \citep{2018PhRvD..98e5032J,2021arXiv210405727C}.

Since ALPs are massive, they arrive on Earth later than photons. The time delay is written as
\begin{equation}
    t(E_\gamma)=\frac{L_1}{\beta}+L_2-d_\mathrm{SN},
\end{equation}
where $E_\gamma=E_a/2$ is the photon energy. Because $L_1$, $L_2$, and $\beta$ {are} dependent on energy, the decay photons with different energies arrive on Earth at different times. {In the region of the parameter space discussed in this study, $L_1\approx\lambda_{a\rightarrow\gamma\gamma}$ is always shorter than $d_\mathrm{SN}$.}

\begin{figure}
 \begin{center}
  \includegraphics[width=80mm]{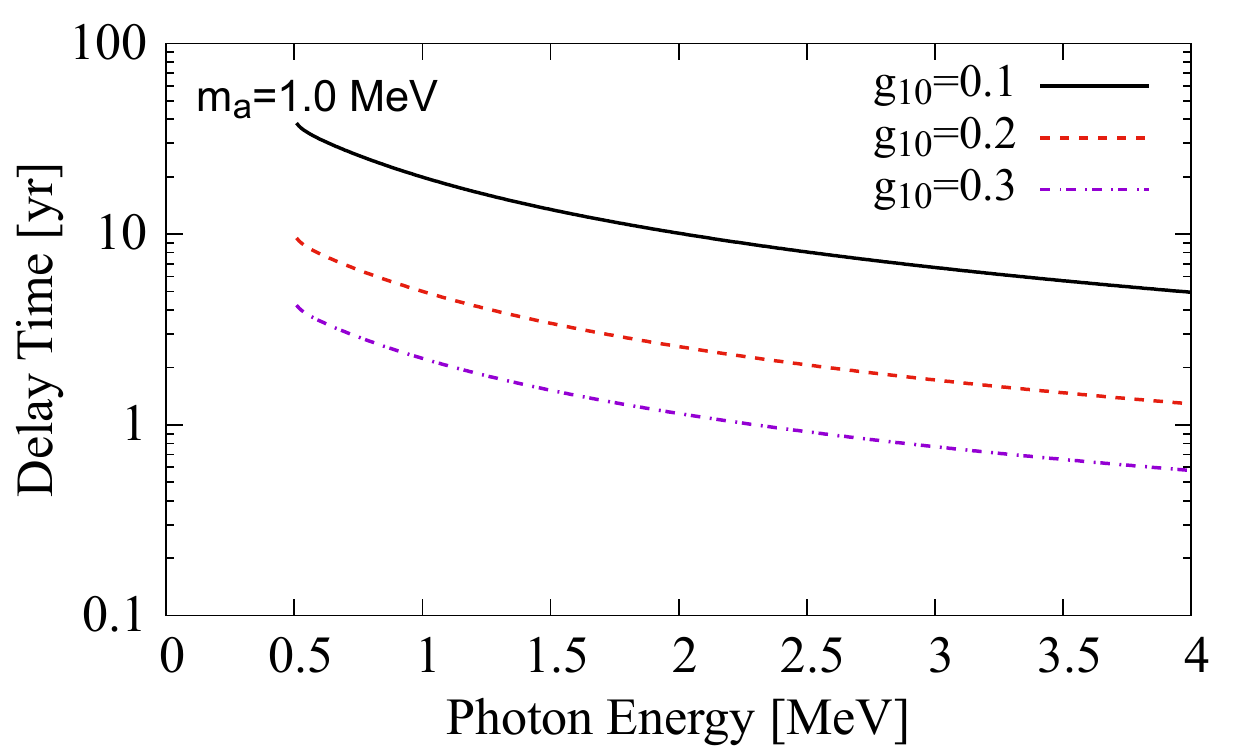}
 \end{center}
 \caption{The delay time $t(E_\gamma)$ of decay photons  observed on Earth. The ALP mass is fixed to $m_a=1$ MeV. 
 }\label{dt}
\end{figure}

Figure \ref{dt} shows the time delay $t$ in the case of $m_a=1$ MeV as a function of  $E_\gamma=E_a/2$. 
It is seen that more energetic photons arrive earlier regardless of $m_a$ because energetic ALPs run faster. 
The delay time becomes shorter when $g_{10}$ is larger since the MFP is shorter when $g_{10}$ is larger and hence the signal runs at the speed of light longer. The low-energy cutoff in the plot is the lowest possible photon energy which is equivalent to $m_a/2$.

Given the delay time of decay photons, it is possible to estimate the energy flux observed on Earth which is defined as
\begin{eqnarray}
    F(t)=E_\gamma(t)\frac{dn_\gamma(t)}{dt},
\end{eqnarray}
where $dn_\gamma/dt$ is the number of photons with the energy $E_\gamma$ per a unit area and time. The photon number is calculated as \citep{2021arXiv210405727C}
\begin{eqnarray}
    \frac{dn_\gamma}{dt}=\frac{2}{4\pi d_\mathrm{SN}^2} \frac{dn_a}{dE_a}\left(\frac{dt}{dE_a}\right)^{-1}e^{-\frac{R_\mathrm{SN}}{\lambda_{a\rightarrow\gamma\gamma}(E_a)}}\nonumber\\
    \times\left(1-e^{-\frac{d_\mathrm{SN}}{\lambda_{a\rightarrow\gamma\gamma}(E_a)}}\right).
\label{na}
\end{eqnarray}
The overall factor 2 corresponds to two decay photons that are produced by an ALP. {The factor $dn_a/dE_a$ is the spectrum of ALPs emitted from the SN and $dt/dE_a$ is the energy derivative of the delay time.} The factor $\exp(-R_\mathrm{SN}/\lambda_{a\rightarrow\gamma\gamma}(E_a))$ is the survival probability of ALPs inside the SN, where $R_\mathrm{SN}=3\times10^3$ km is assumed on the basis of the mass-radius relation of massive white dwarfs  \citep{1961ApJ...134..683H} If an ALP decays in the SN, the decay photon cannot escape to the interstellar space. Another factor $1-\exp(-{d_\mathrm{SN}}/{\lambda_{a\rightarrow\gamma\gamma}(E_a)})$ is the ALP decay probability between the SN and Earth. Since we are focusing on relativistic ALPs, the decay photons are strongly beamed toward the forward direction. We therefore ignore photons that propagate backward with this factor. 

\begin{figure}
 \begin{center}
  \includegraphics[width=80mm]{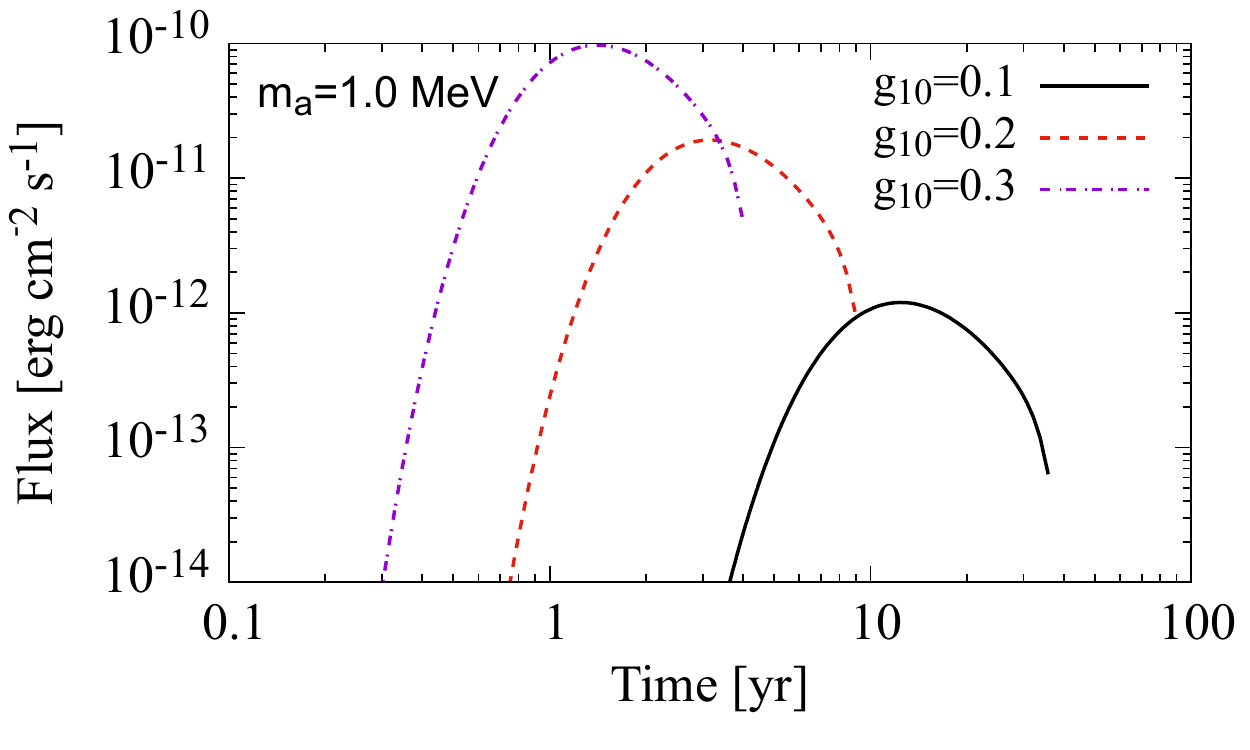}
 \end{center}
 \caption{The flux $F(t)$ of decay photons observed on Earth. The ALP mass is fixed to $m_a=1$ MeV.
 }\label{flux}
\end{figure}

Figure \ref{flux} shows the flux $F$ of decay photons as a function of time after the SN explosion. It is seen that the photon flux is diluted in time and the peak is expected 1-10 years after the explosion. The flux is larger if the ALP-photon coupling is larger because the intrinsic ALP emission from an SN increases as a function of $g_{10}$.  
The arrival time of the peak becomes earlier if $g_{10}$ is larger as we saw in figure \ref{dt}. {The photon flux and the delay time at the peak are summarized in table 1.}
\begin{table*}[t]
    \centering
    \begin{tabular}{c|cccc}
    $g_{10}$&Delay Time [yr]&Peak Flux [erg cm$^{-2}$ s$^{-1}$]&$E_\gamma$ [MeV]&$\theta$ [deg]\\\hline
      0.1   & 12&$1.2\times10^{-12}$&1.6&0.55 \\
    0.2     &3.1&$1.9\times10^{-11}$&1.6&0.14 \\
    0.3&1.4&9.7$\times10^{-11}$&1.6&0.061
    \end{tabular}
    \caption{The summary on decay photons and the angular radius of the halo when the flux reaches its peak. The ALP mass is fixed to 1 MeV.}
    \label{tab:my_label}
\end{table*}
\begin{figure}
 \begin{center}
  \includegraphics[width=80mm]{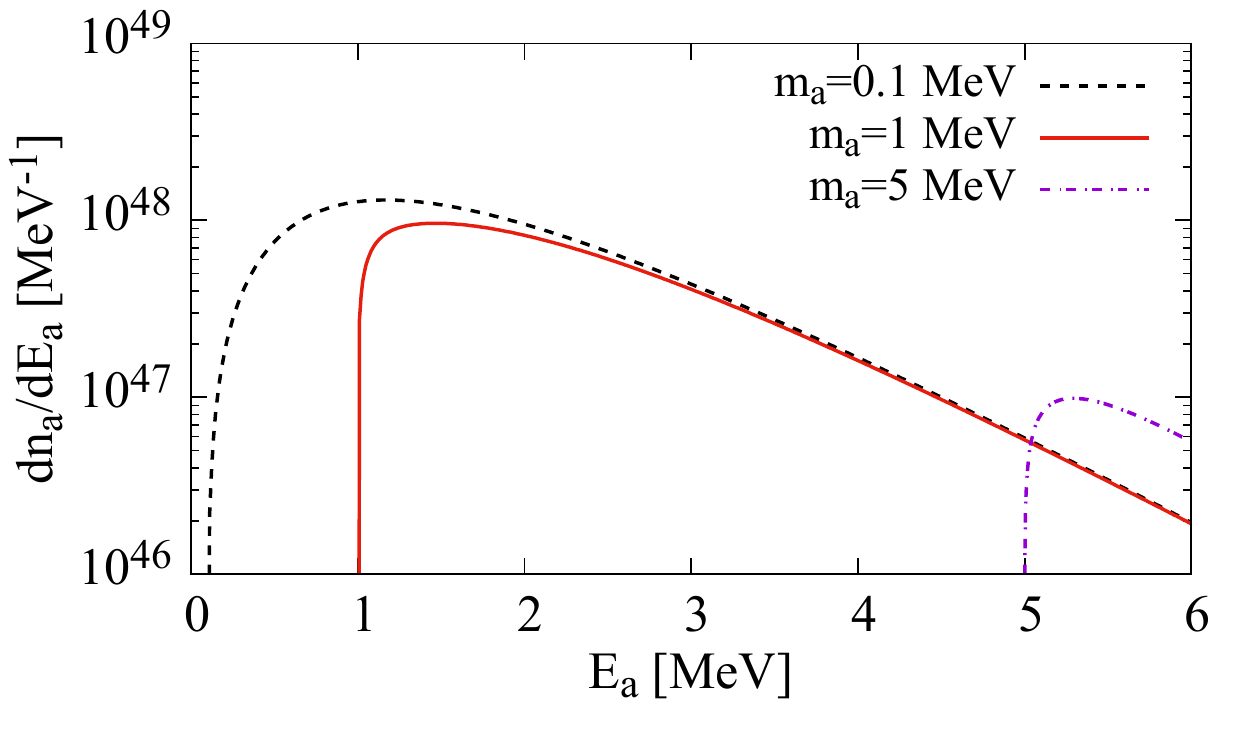}
 \end{center}
 \caption{The spectrum $dn_a/dE_a$ of ALPs emitted from the SN. The coupling constant is fixed to $g_{10}=1$.}\label{dn_dE}
\end{figure}

{Figure \ref{dn_dE} shows the spectrum of ALPs produced in the SN as a function of the ALP energy $E_a$. In the case of $m_a=1$ MeV, the peak in the ALP spectrum is at $E_a\sim1.5$ MeV, while table 1 shows that ALPs that produce the photon flux peak are as energetic as $E_a\sim3.2$ MeV. The difference stems from the time delay. As seen from figure \ref{dt}, low-energy ALPs arrive on Earth later and thus the flux of their decay photons is diluted in time. As a result, the peak in the photon flux is more energetic than the peak in the ALP spectrum.}

{Since the incidence direction of ALPs does not coincide with the line of sight between the SN and Earth, the decay photons would form a halo around the SN when they are observed on Earth. When $L_1$ is close to $d_\mathrm{SN}$, the size of the halo can be so large that it exceeds a finite angular acceptance of detectors \citep{2018PhRvD..98e5032J}. When $m_a$ is fixed, the halo is larger if $g_{10}$ is smaller because $L_1$ is longer. It is easily shown from figure \ref{sche} that the angular radius $\theta$ of the halo is written as
\begin{equation}
    \sin\theta=\frac{L_1}{d_\mathrm{SN}}\sin\alpha.
\end{equation}
The values of $\theta$ at the peak of the photon flux are shown in table 1. It is seen that the halo is larger than Moon if $g_{10}\lesssim0.1$. Such a large halo might make observations of decay photons challenging.}

The photon flux $\sim1$ year after the SN event is as high as $\sim10^{-11}$ erg cm$^{-2}$ s$^{-1}$ if  $g_{10}\gtrsim0.3$ and $m_a=1$ MeV. We can see from figure \ref{dt} that $E_\gamma=O(\mathrm{MeV})$ at the peak of the flux.  The sensitivity of instruments for MeV photons on current $\gamma$-ray satellites is not enough to detect the signal because the instrumental background prevents one from improving it.  \citep{2004NewAR..48..193S}. However, next-generation instruments including the Electron-Tracking Compton Camera (ETCC: \cite{2015ApJ...810...28T}) and All-sky Medium Energy Gamma-ray Observatory (AMEGO: \cite{2019BAAS...51g.245M}) plan to achieve sensitivities as high as $\sim10^{-12}-10^{-11}$ erg cm$^{-2}$ s$^{-1}$. Observing MeV $\gamma$-rays from a nearby SN Ia with the next-generation instruments may give an independent constraint on the ALP parameters.

A recent nearby core-collapse SN, namely SN 1987A, has provided a stringent constraint on ALPs with $m_a\sim10$ keV-100 MeV \citep{2011JCAP...01..015G,2018PhRvD..98e5032J}. \citet{2018PhRvD..98e5032J} used an upper limit on the $\gamma$-ray fluence obtained by the Solar Maximum Mission \citep{1989PhRvL..62..505C} to constrain $g_{10}$. They concluded that $g_{10}\lesssim0.2$ for ALPs with $m_a=1$ MeV. If the ALP parameters are on this limit, the $\gamma$-ray flux from a nearby SN Ia can be as high as $10^{-12}-10^{-11}$ erg cm$^{-2}$ s$^{-1}$, as can be seen from figure \ref{flux}. Since this flux may be observable with next-generation instruments, SNe Ia can potentially provide an independent constraint which is comparable to the constraints obtained with core-collapse SNe. 
\section{Summary and Discussion}
In this study, I calculated the emission of heavy ALPs from a nearby SN Ia for the first time. It is found that the ALP luminosity reaches $L_a\sim 10^{43}g_{10}^2$ erg s$^{-1}$ for $m_a\lesssim1$ MeV. The emission is suppressed for heavier ALPs because thermal photons are not energetic enough. I then estimated the flux of MeV decay photons from an SN Ia that is located 1 kpc away. The photons observed on Earth are diluted in time: energetic photons arrive earlier. The flux can be as high as $10^{-12}-10^{-11}$ erg cm$^{-2}$ s$^{-1}$ if the ALP parameters are on the SN 1987A limit \citep{2018PhRvD..98e5032J}. Next-generation MeV $\gamma$-ray telescopes including ETCC and AMEGO may provide an independent constraint on $m_a$ and $g_{10}$ which can be as stringent as the SN 1987A limit if an SN Ia happens $\sim1$ kpc away or closer.

In my calculation, a near-Chandrasekhar mass model is adopted \citep{1984ApJ...286..644N}. However, the nature of SN Ia progenitors is still under debate. Recently, sub-Chandrasekhar mass progenitor models have been developed as well  \citep{2010A&A...514A..53F,2011ApJ...734...38W,2018ApJ...854...52S,2020ApJ...888...80L}. Since the central density in a sub-Chandrasekhar mass WD can be as low as $\sim10^7$ g cm$^{-3}$, the ALP emissivity can be significantly affected. It is hence desirable to perform a systematic study that focuses on the model dependence.

{In this study, SNe are not assumed to be magnetized.  However, magnetic fields in a white dwarf can be as strong as $\sim100$ MG \citep{2003ApJ...595.1101S}. In such a case, the flux of decay photons can be significantly higher than the unmagnetized case because of two reasons. One is that plasmon-ALP conversion in plasma by external magnetic fields can enhance the intrinsic ALP production in a white dwarf \citep{1998PhRvD..58e5008M,2020PhRvD.101l3004C,2021arXiv210405727C}. Another reason is that magnetic fields can convert ALPs into photons to enhance the photon flux \citep{2019PhRvL.123f1104D}. Systematic studies on the effects of magnetic fields are desirable.} 

It has been reported that neutrinos from a nearby SN Ia may be observable and provide information on the explosion mechanism \citep{2007PASJ...59L..57K,2011A&A...529A.156O,2016PhRvD..94b5026W}. {The predicted peak luminosity of neutrinos reaches $\sim10^{50}$ erg s$^{-1}$, which is much higher than the ALP luminosity. The ALP emission is thus not likely to affect the neutrino emission.} Multi-messenger astronomy of SNe Ia would shed light on both of the nature of SNe and fundamental physics.


\begin{ack}
The author thanks Kei Kotake and Tomoya Takiwaki for fruitful discussions. This work is supported by Research Institute of Stellar Explosive Phenomena at Fukuoka
University.
\end{ack}

\end{document}